\documentstyle[prb,aps,epsfig,multicol]{revtex}

\begin{document}
\title{Electrons and Phonons in YbC$_{6}$}
\author{I.I. Mazin}
\address{Center for Computational Materials Science,\\
Naval Research Laboratory, Washington, DC 20375}
\author{S.L. Molodtsov}
\address{Institut f\"ur Festk\"orperphysik, TU Dresden, D-01062 
Dresden, Germany}
\date{\today }
\maketitle

\begin{abstract}
The electronic structure and selected zone-center phonons in Yb graphite intercalation compound 
(YbC$_{6}$) are investigated {theoretically} using density functional calculations and LDA+U 
approximation for Coulomb correlations in the $f$ shell, and experimentally using angle-resolved 
photoemission. We find that both in LDA and LDA+U approach the Yb $f$ states are fully occupied 
providing no evidence for mixed-valent behavior. The obtained theoretical results are in good 
agreement with photoemission experiments. The 4$f$ states are considerably hybridized both with 
the Yb 5$d$ and C 2$p$ states resulting in a substantial admixture of Yb $f$ at the Fermi level. 
Soft Yb phonons, given a noticeable presence of the Yb states at the Fermi level, are probably 
responsible for the superconductivity recently reported in YbC$_{6}$.
\end{abstract}


\begin{multicols}{2}


For a long time graphite intercalation compounds (GICs) have been attracting substantial attention 
due to their layered quasi-two dimensional (2D) structure and the resulting large anisotropy of 
their electric and electronic properties.\cite{interc1,interc2} Some of GICs are known to be 
superconducting at low temperatures.\cite{superc} A renewed interest in superconducting phenomena 
in intercalated graphites is due to the recent discoveries of superconductivity in MgB$_{2}$ and 
B$_{x}$C, electronically related to graphite.\cite{mgb2} An interesting example of graphite 
intercalation compounds is YbC$_{6}$. First, it is one of the only four known bulk intercalations 
with 4$f$ metals: YbC$_{6},$ EuC$_{6}$, SmC$_{6}$, and TmC$_{6}$ \cite{binterc} which may indicate 
that $f$ electrons play certain role in bonding in these compounds. Likewise, YC$_{6}$ has not 
been synthesized so far, although Y behaves similar to lanthanides in systems with localized $f$ 
electrons (cf. high-T$_c$ cuprates). Last but not least, very recently superconductivity was 
discovered in YbC$_{6}$.\cite{unpb}

Yb is known to form mixed-valent compounds,\cite{mv} so one may think mixed-valent physics is 
operative in YbC$_{6}$ as well. Therefore, it is of interest to investigate the electronic 
structure of YbC$_{6}$ numerically and experimentally. Should a good agreement between the theory 
and experiment be established, the former can be used to gain some insight into possible mechanism 
for superconductivity. Experimentally, Yb valence in the above compound can be obtained by 
recording the Yb 4$f$ photoemission (PE) spectra. Due to the large Coulomb-correlation energy, the 
energy positions of the PE signals for 4$f$ configurations with different electron occupations are 
shifted by several eV with respect to each other. Therefore, the contributions to the PE from 
different 4$f$ configurations can easily be discriminated.\cite{wieling} From the intensity ratio 
of these contributions the information about the valence can be derived.

In this paper we present {\it ab initio} calculations of the electronic structure of YbC$_{6}$ 
using both local density approximation (LDA) and LDA+U approach, which accounts for Coulomb 
correlations inside the $f$ shell. The calculated energies for the Yb 4$f$ states are in good 
agreement with the results of photoemission experiments giving no evidence for the mixed-valent 
behavior. We also report first principles calculations of selected zone center phonons. We find 
soft Yb-derived phonons to be likely responsible for the superconductivity reported in YbC$_{6}$.

\begin{figure}[tbp]
\centerline{\epsfig{file=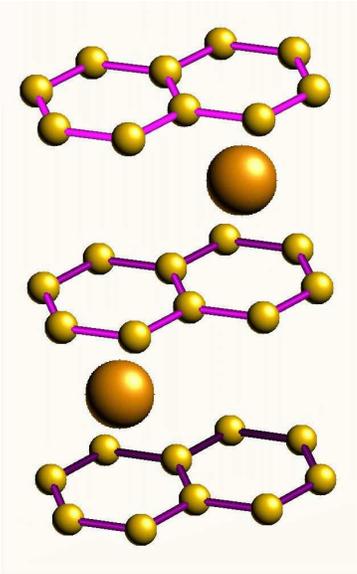,width=0.55\linewidth,angle=0,clip=}} \vspace{0.3cm} 
\caption{Crystal structure of YbC$_6$. Large spheres denote Yb and small spheres denote C. (color 
online)} \label{str}
\end{figure}

For the calculations, we used experimental crystal structure $P6/mmc,$ with C occupying 12$i$ and 
Yb 2$d$ positions, with the lattice parameters $a=4.32$ \AA\ and $c=9.1$ \AA\ (Fig. \ref{str}). A 
full potential linear augmented plane wave method (LAPW) was used~\cite{WIEN} with the following 
setup: The APW sphere radii were taken as 2.5 and 1.3 bohr, the cutoff parameter $RK_{\max }=7,$ 
and local orbitals were used for Yb $s$ and $p$ and for C $s$, to reduce the linearization error, 
and to improve convergence in $RK_{\max }$. Linear tetrahedron method was employed for the 
Brillouin zone (BZ) integration, with the {\bf k} mesh up to 11$\times 11\times 4$ divisions. 
Spin-orbit coupling was included on the second variational basis.\cite{WIEN} The Ceperley-Alder 
exchange-correlation potential was used in the LDA part  of the calculations. Finally, Hubbard 
correlations in the $f$ shell were taken into account using the fully localized~\cite{Pet} version 
of the LDA+U formalism, with the parameters $U=0.4$ Ry and $J=0.05$ Ry, estimated by modifying 
occupation numbers in the quasiatomic loop in a LMTO (linear muffin-tin orbital) program. 

The Yb-GIC samples for the PE experiments were prepared {\it in situ} by thermally driven surface 
reaction of deposited Yb overlayers with graphite (0001) substrate as described 
elsewhere.\cite{prep} The measurements were performed using synchrotron radiation from various 
beamlines at BESSY/Berlin electron storage ring. Angle-resolved photoemission spectra (ARPES) were 
taken with 33 eV photon energy using rotative hemispherical electron energy analyzers (VSW-ARIES) 
tuned to an energy resolution 150 meV (full width at half maximum, FWHM) and an angular resolution 
of 1$^{\circ}$. Basic pressure during measurements was always better than 1$\times$10$^{-10}$ mbar.

\begin{figure}[tbp]
\centerline{\epsfig{file=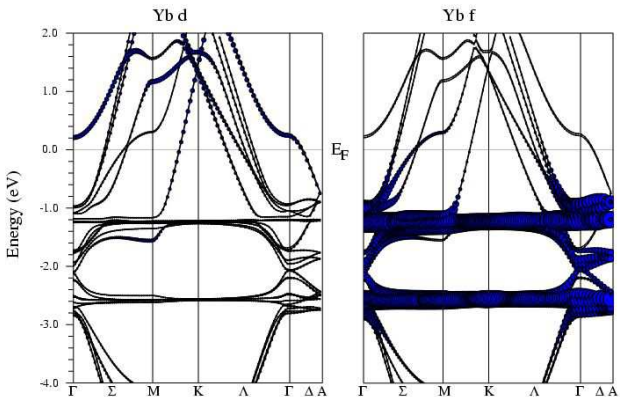,width=1.00\linewidth,angle=0,clip=}} \vspace{0.3cm} 
\caption{LDA+U band structure of YbC$_6$. The left panel shows the partial Yb-$d$ character, and 
the right panel Yb-$f$ character. (color online) } \label{BS}
\end{figure}

The band structure calculated within the LDA+U scheme is shown in Fig. \ref{BS}. Probably the most 
intriguing observation is that all $f$ states appear to be fully occupied, giving evidence to 
firmly divalent Yb. This seems to be related to the fact that straight LDA calculation (not 
shown), without any account of intra-$f$-shell Hubbard-type correlations, still place all $f$ 
bands {below} the Fermi level, although, of course, much higher {in energy} than in LDA+U.

\begin{figure}[tbp]
\centerline{\epsfig{file=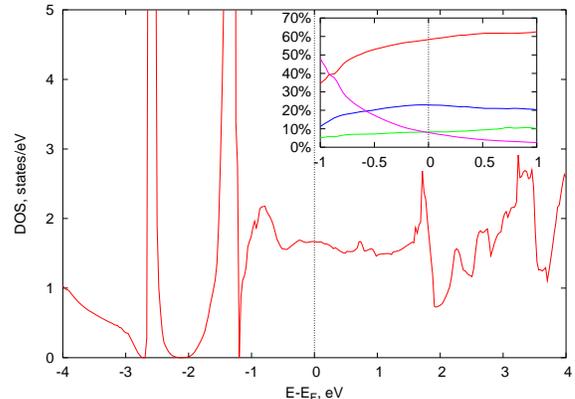,width=0.90\linewidth,angle=0,clip=}} \caption{Electronic density 
of states of YbC$_6$. The main graph shows the total DOS on a per-formula-unit basis and the 
insert shows, in the order from the lowest to highest DOS at $E-E_F=1$ eV, the partial 
contribution of Yb $f$, Yb $d$, C $p$ character inside the corresponding APW spheres and the 
contribution of interstitial states. (color online)} \label{DOS}
\end{figure}

The spin-orbit interaction splits the $f$ states into two manifolds, located 1.10 and 2.30 eV 
below the Fermi level (E$_F$). The corresponding total electronic density of states (DOS) with two 
strong peaks originating in the 4$f$ states is demonstrated {in Fig. \ref{DOS}}. Calculated 
positions of the 4$f$ states in lanthanide systems depend very much on specific choice of $U$ and 
$J$ values, which are often used as fitting parameters to achieve better correspondence with the 
experimental data. It is very important to note that in the present study the $U$ and $J$ values 
were estimated from the first principles and no attempts were done to fit or tune this parameters. 

\begin{figure}[tbp]
\centerline{\epsfig{file=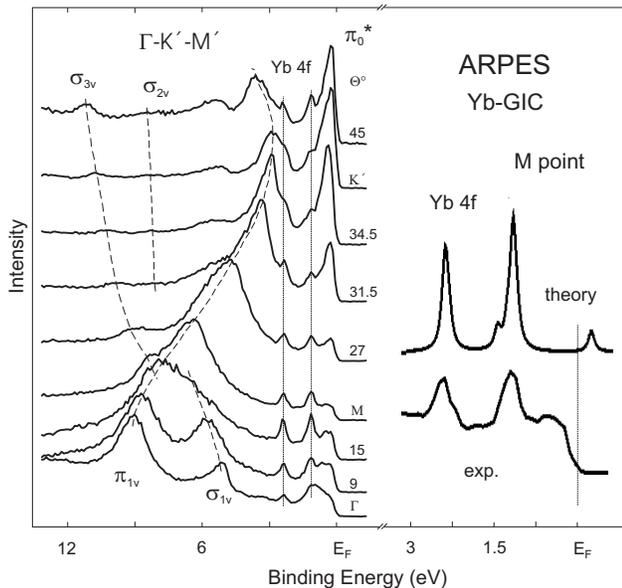,width=0.95\linewidth,angle=0,clip=}} \caption{(Left) 
angle-resolved PE spectra of the Yb-GIC taken along the $\Gamma$-$K$'-$M$' direction in the BZ of 
graphite ($\Gamma$-$M$-$\Gamma$-$M$ in the BZ of the Yb-GIC) at different polar angles $\Theta$. 
(Right) Comparison between the local DOS in the vicinity of the $M$ point and the angle-resolved 
PE spectrum sampling this region. } \label{SM}
\end{figure}

Most importantly, the qualitative prediction of purely divalent Yb in Yb-GIC is unambiguously 
supported by the PE data. Fig. \ref{SM} depicts a series of angle-resolved PE spectra taken along 
the $\Gamma$-$K$'-$M$' direction in the BZ of graphite. All four characteristic occupied bands of 
graphite ($\sigma_{1v}$, $\sigma_{2v}$, $\sigma_{3v}$, and $\pi_{1v}$) are clearly seen in the 
figure pointing to the high-quality intercalation compound under consideration.\cite{prep} The 
Fermi-energy peak appearing in the region close to the $K$' point stems from the bottom of the 
$\pi_{0}^\ast$ band of graphite, which becomes occupied in the GIC due to charge transfer of 
electrons from the intercalant atoms. The none-dispersive divalent contributions into the PE 
signal of the Yb 4$f$ electrons, which are found at 1.14 eV and 2.41 eV binding energies (BEs), 
are marked by two thin lines through the spectra. Would the divalent 4$f$ component of the 
final-state PE multiplet be observed at the Fermi energy, it will show that the final-state 
multiplet is energetically degenerate with the ground state that will provide a reason for 
homogeneous mixed-valent behavior of the system. In the present case of the Yb-GIC, the divalent 
signal is observed too far from E$_F$ to anticipate the mixed-valent properties of the compound. 
Correspondingly, no traces of a trivalent signal in the region of 5 to 12 eV BE,\cite{wieling} 
which one would expect for a mixed-valent system, are seen in the photoemission spectra. 

The experimentally obtained binding energies of the divalent 4$f$ contributions are in excellent 
agreement with the theoretical values. Not only energies, but also lineshapes of the DOS and PE 
spectra are in agreement with each other. In the right panel of  Fig. \ref{SM} we compare the 
angle-resolved PE spectrum sampling electronic states in the region of the $M$ point in the BZ of 
the Yb-GIC with the local DOS (spectral function) calculated around this point. The calculated 
local DOS was broadened with a Lorentzian to account for finite lifetime effects. Good 
correspondence between the theoretical and experimental data is evident in the figure. In fact, 
the agreement is better than one expects given the approximations used. Nevertheless, even though 
such level of accuracy is likely fortuitous, this clearly indicates that the underlying physics is 
correctly described by the calculations, and, importantly, that the derived value of $U$ and $J$ 
are in the right ballpark. 

\begin{figure}[tbp]
\centerline{\epsfig{file=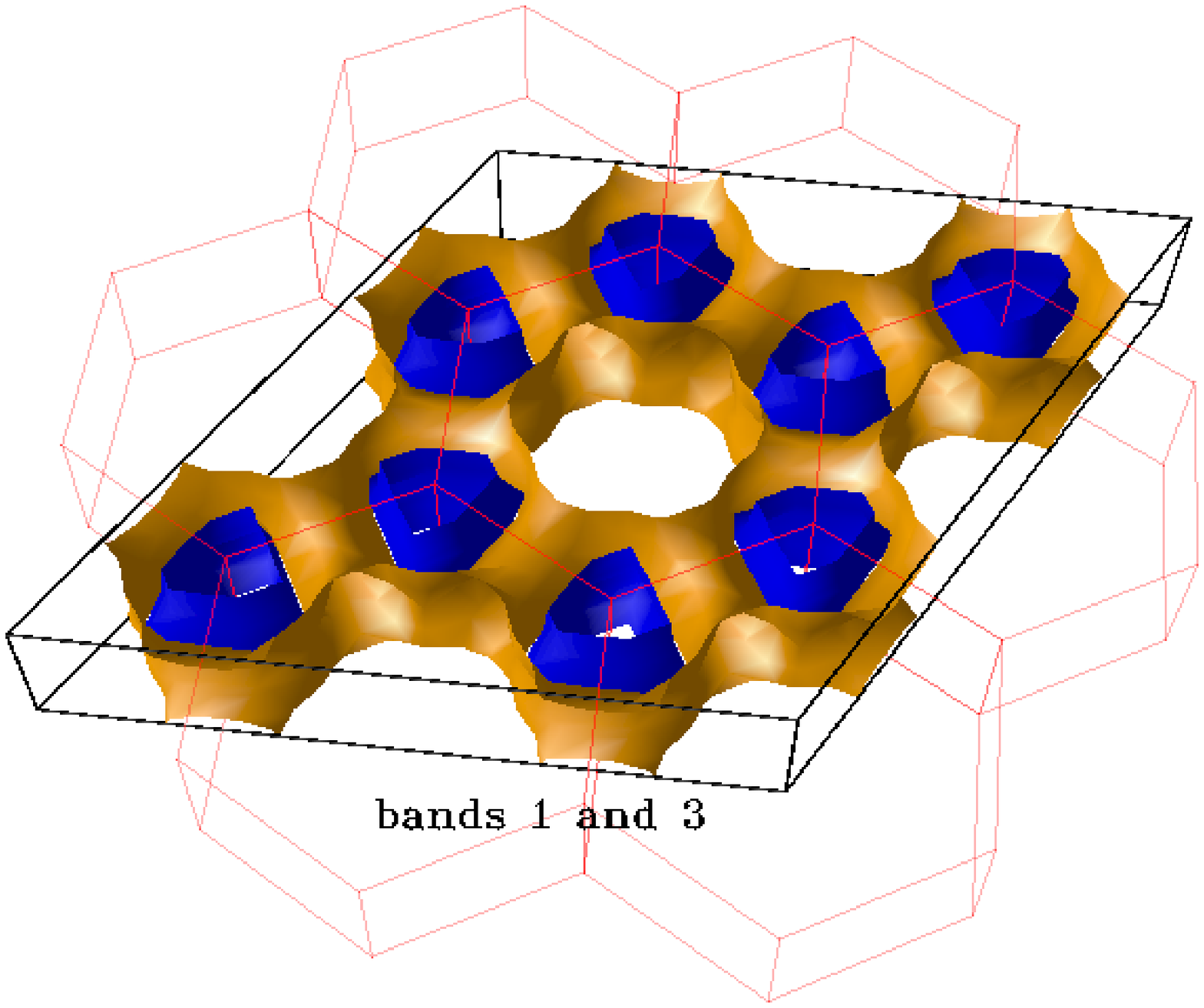,width=0.75\linewidth,angle=0,clip=}} 
\centerline{\epsfig{file=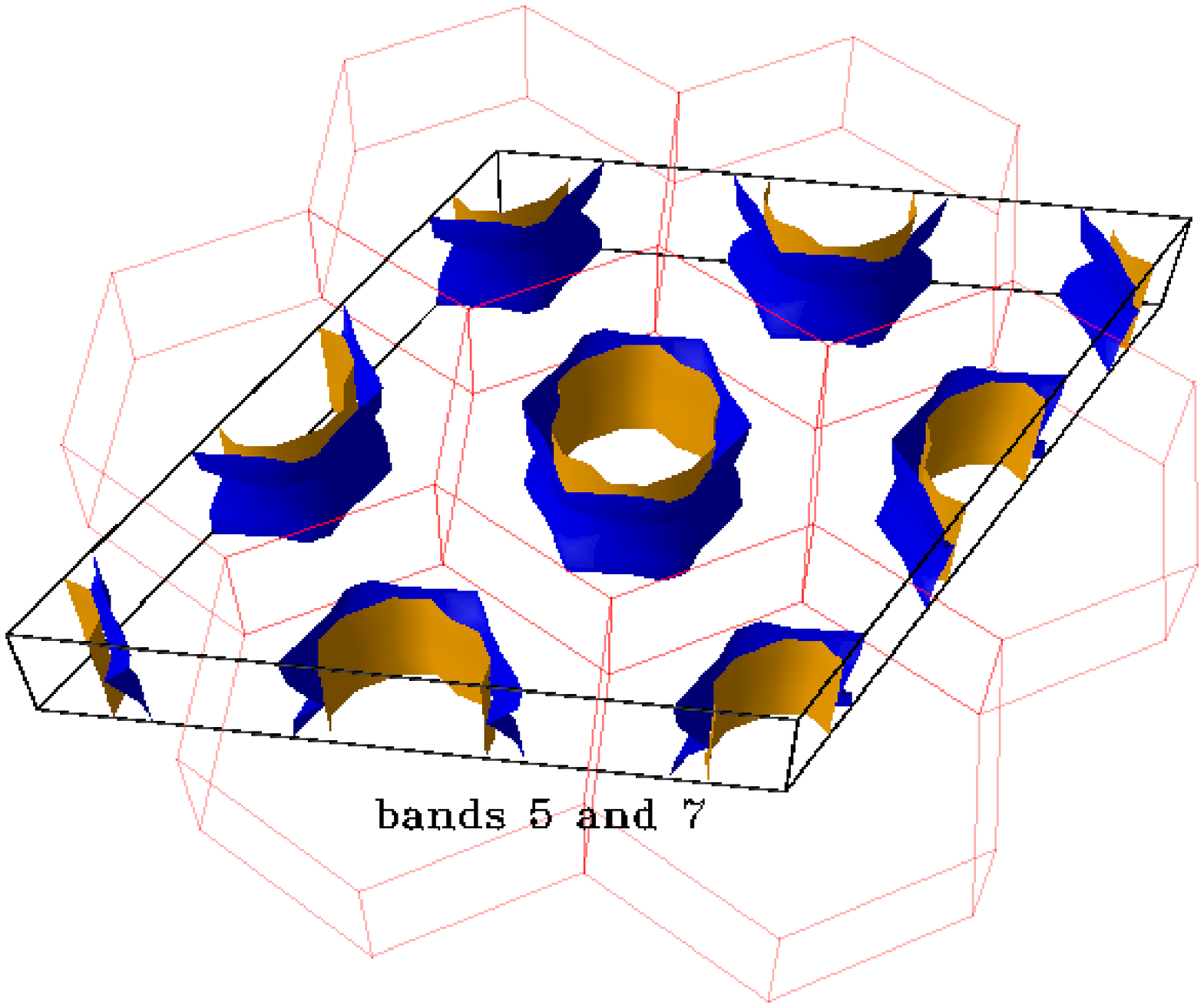,width=0.75\linewidth,angle=0,clip=}} 
\centerline{\epsfig{file=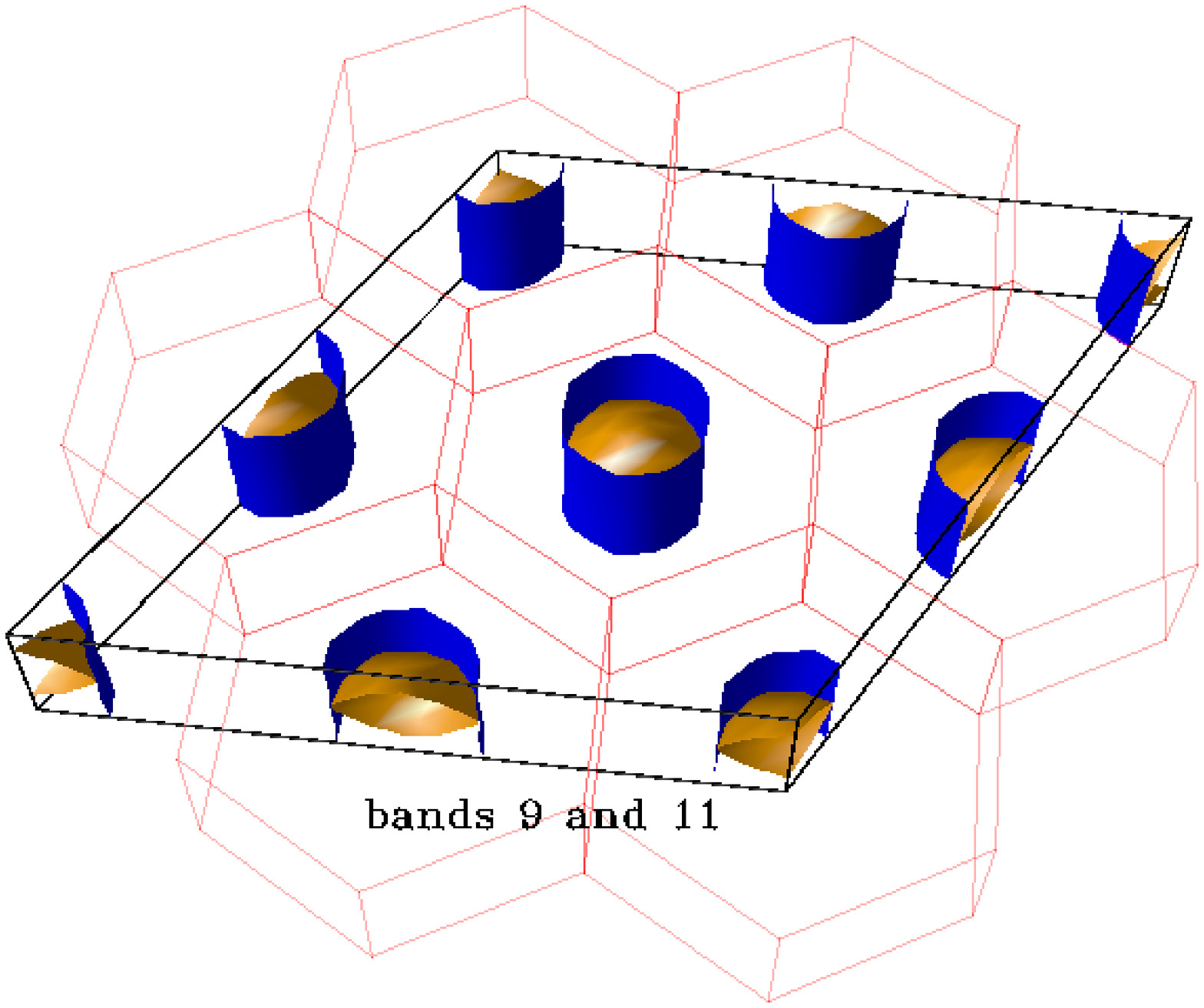,width=0.75\linewidth,angle=0,clip=}} \caption{LDA+U Fermi 
surface of YbC$_6$. The band with the lower index is blue (darker in grey scale). (color online)} 
\label{FS}
\end{figure}

We shall now discuss the calculated electronic structure in more detail. Even in the vicinity of 
the Fermi level, the bands of YbC$_6$ are qualitatively different from that of YC$_{6}$ (Fig. 
\ref{BS}), indicating that the hybridization of the $f$ states with the itinerant electrons should 
play an important role in transport properties in this material. There are 12 bands crossing the 
Fermi level. However, spin-orbit splitting at E$_F$ is very small, so that one can safely speak 
about 6 distinct Fermi surfaces, corresponding to bands 1, 3, 5, 7, 9, and 11, shown in Fig. 
\ref{FS}. An easy way to make the correspondence between the Fermi surfaces and the bands plotted 
in Fig. \ref{BS} is to count them from $K$ to $\Gamma$: the only band crossing the Fermi level 
between $K$ and $M$ is band 1(2), while the state at approximately 0.2 eV at $\Gamma$ is band 
11(12). Bands 1, 5, 7, and 9 are quasi 2D, while bands 3 and 11 have considerable dispersion along 
the $z$ direction. It is instructive to look at the element specific character of these bands. 
Band 11, forming elliptical pockets around $\Gamma$, has more Yb character than C, and namely Yb 
$d_{z^{2}-r^{2}}$ character. Not surprisingly, this is the band that is most dispersive in the $z$ 
direction. Bands 5 through 8 have predominantly C $p_{z}$ origin (band 5 with a considerable 
hybridization with Yb $f$), while bands 1 and 3 show only small contributions within either C or 
Yb muffin-tin spheres, but rather in the interstitial areas, thus being mostly free-electron like.

While at the Fermi level most of the DOS (inset in Fig. \ref{DOS}) comes from the extended states in the 
interstitial region, there is a noticeable share of the Yb $f$ states, which rapidly grows with 
increasing BE. Contribution of the Yb $d$ states (mostly $d_{z^{2}-r^{2}}$ symmetry) is also 
considerable. This indicates that (i) Yb phonons may play a role in superconductivity and (ii) 
substituting Yb by other divalent elements, like Ca, may change superconducting properties 
substantially, or destroy superconductivity at all. The total density of states at the Fermi level 
is $N(E_{F})=1.7$ states/(eV$\cdot $formula). The average Fermi velocity in- and out-of-plane, 
respectively, is 4.8$\times 10^{7}$ and 2.7$\times 10^{7}$ cm/sec. The corresponding components of 
the plasma frequency are 6.2 and 3.6 eV, implying, in the constant scattering rate approximation, 
a resistivity anisotropy of about 3.

The large unit cell of YbC$_{6}$ allows for multiple zone-center modes, which in principle can be 
computed by the frozen phonon method. 24 modes corresponding to in-plane motion of C are expected 
to be very hard, since C $p_{x,y}$ orbitals are removed from the Fermi level (as opposed, for 
instance, to MgB$_{2})$. Indeed, we calculated the frequency of the fully symmetric $A_{1g}$ mode 
to be 1540 cm$^{-1}$. Most of the 12 out-of plane C modes, by symmetry, have zero deformation 
potential at $\Gamma$ and thus do not couple with electrons, while the $E_{g}$ mode lowers the 
symmetry to triclinic and is prohibitively difficult to compute accurately. Thus, we are left with 
the 6 Yb modes, of which the three even modes correspond to a double degenerate $E_{2g}$ phonon 
involving displacements of Yb parallel to the graphite plane, and a $B_{1g}$ one, corresponding to 
Yb displacements along $c$. These modes are rather soft; our calculations place them at 77 and 153 
cm$^{-1}$, respectively. Unfortunately, the electron-phonon coupling exactly at the zone center is 
strongly suppressed by symmetry for these three modes (if the alternating Yb planes were not 
shifted with respect to each other, cf. Fig. \ref{str}, it would be forbidden, so in the actual 
structure it is suppressed to the extent of the weakness of Yb-Yb interaction. However, their low 
frequency and considerable presence of Yb character at the Fermi level suggest that coupling of 
these and similar modes, when integrated over the Brillouin zone, may be sufficient to explain 
superconductivity in YbC$_{6}$.

Work at the Naval Research Laboratory was supported by the Office of the Naval Research. 
Experimental part was supported by the Deutsche Forschungsgemeinschaft, SFB 463 (TP B16).

\end{multicols}

\end{document}